\documentstyle[preprint,eqsecnum,aps,psfig]{revtex}
\begin{document}
\bibliographystyle{unsrt}
\title{Bader's interatomic surface and Bohmian mechanics.} 
\author{L.Delle Site\\ 
Max-Planck-Institute for Polymer Research\\
Ackermannweg 10, D 55021 Mainz\\
Germany }
\date{\today}
\maketitle

\begin{abstract}
A Thomas-Fermi statistical analysis of Bader's interatomic surface
developed in a previous work \cite{luigiletter} is here extended by considering
exchange effects and electron density's inhomogeneity at basic level
via Thomas-Fermi-Dirac-Weizsacker model.
The results obtained show interesting connections with
bohmian mechanics and lead to a statistical interpretation of
the chemical properties of condensed systems at atomistic level.
\end{abstract}    
\section{Introduction}
In a previous work \cite{luigiletter} we analyzed the statistical
meaning of Bader's interatomic surface (or zero flux surface of the
electronic density gradient), i.e.
\begin{equation} 
{\bf \nabla}\rho({\bf r})\cdot {\bf n}=0,\forall {\bf r} \in S({\bf
  r})
\label{surface}
\end{equation}
(where ${\bf n}$ is the unit vector perpendicular to $S({\bf
r})$
at each point ${\bf r}$ of the surface $S$).

We used the approximation of
Thomas-Fermi for the electron density and related properties and  
showed that the partitioning 
of the whole system in subsystems which are in mutual statistical
equilibrium is consistent with Bader's partitioning of atoms in
molecules. The Theory of {\it Atoms in Molecules} is well known, thus
we do not review it and suggest the reader to consult the following
references \cite{bader1,bader2,bader3,bader4,bader5,bader6}, while for 
technical applications see for example
\cite{luigi3,luigi4,luigi5,luigi6,popelier1,popelier2,xanteas}. The
central point of our previous work was to define 
the electronic chemical potential within the
approximation of the simple Thomas-Fermi model and then consider two
cases, electron as non interacting particles, and electrons as
classical interacting particles. Next, by imposing the condition of
statistical equilibrium at each point of the surface of separation between two
subsystems, which means the gradient of the electronic chemical potential
at the surface to be zero along the crossing direction (normal to the
surface), we could comment on the 
consistency between Bader's definition and the statistical definition 
of atomic subsystem in a multiatomic system.
In this work we extend the previous analysis to 
a more sophisticated Thomas-Fermi model
which takes into account exchange energy (Dirac) and the inhomogeneity 
of the electron density, i.e. the Weizsacker correction to the
Thomas-Fermi kinetic energy; the resulting model is known as
Thomas-Fermi-Dirac-Weizsacker (TFDW) (see pg 127-136 of \cite{DFT}). We apply the procedure developed 
in our previous work and find some interesting connections with bohmian
mechanics (for details about bohmian mechanics see for example \cite{bohm1,bohm2,bohm3} and references therein). 
This is an interesting result since an attempt to
interpret Bader's theory in terms of bohmian mechanics has already been
done for the part relative to the chemical bond \cite{bohm4,bohm5,bohm6}; 
our current results
can be integrated with those and lead to a simple 
interpretation of some chemical concepts in a statistical context with
evident advantages for molecular modeling and its applications.
\section{Thomas-Fermi-Dirac-Weizsacker model}
The introduction in the simple Thomas-Fermi energy functional 
\begin{equation}
E_{TF}=C\int_{V}\rho({\bf
  r})^{5/3}-\sum_{i=1,M}Z_{i}\int_{V}\frac{\rho({\bf r})}{|{\bf
    r}-{\bf R}_{i}|}d{\bf r}+\frac{1}{2}\int\int_{V}\frac{\rho({\bf
    r})\rho({\bf r'})}{|{\bf r}-{\bf r}'|} d{\bf r}d{\bf r}'
\label{tote}
\end{equation}
(where C is a proper dimensional constant, $V$ is the total volume of
the system, the first term represents the kinetic energy within the 
Thomas-Fermi approximation, the second term describes the total interaction
between electrons and nuclei ($Z_{i},{\bf R}_{i}$ 
represents the charge and position of the $i$-th nucleus, while
$i=1,M$))
of the exchange energy term $-C_{x}\int\rho({\bf r})^{4/3}d{\bf r}$ where
$C_{x}$ is the appropriate dimensional constant and the Weizsacker
correction term to the kinetic energy
$\frac{\hbar^{2}}{8m}\int\frac{|\nabla\rho({\bf r})|^{2}}{\rho({\bf r})}d{\bf
  r}$,
leads to the, so called, TFDW energy functional:
\begin{equation}
 \begin{array}{ll}
E_{TFDW}=C\int_{V}\rho({\bf
  r})^{5/3}-\sum_{i=1,M}Z_{i}\int_{V}\frac{\rho({\bf r})}{|{\bf
    r}-{\bf R}_{i}|}d{\bf r}+\frac{1}{2}\int\int_{V}\frac{\rho({\bf
    r})\rho({\bf r'})}{|{\bf r}-{\bf r}'|} d{\bf r}d{\bf
  r}'\\-C_{x}\int\rho({\bf r})^{4/3}d{\bf r}+\frac{\hbar^{2}}{8m}\int\frac{|\nabla\rho({\bf r})|^{2}}{\rho({\bf r})}d{\bf
  r}
\end{array}.
\label{tfdw}
\end{equation}
Within this approximation the electronic chemical potential $\mu({\bf r})$ is derived via a
variational procedure as
\begin{equation}
\mu({\bf r})=\frac{\delta E_{tot}[\rho({\bf r})]}{\delta\rho({\bf r})}.
\end{equation} 
and this leads to the following relation:
\begin{equation}
\begin{array}{ll}
\mu({\bf r})=\frac{5}{3}C\rho({\bf
  r})^{2/3}-\sum_{i=1,M}
\frac{Z_{i}}{|{\bf r}-{\bf R}_{i}|}+
\frac{1}{2}\int_{V}\frac{\rho({\bf r'})}{|{\bf r}-{\bf r}'|} d{\bf r}'\\
-\frac{4}{3}C_{x}\rho({\bf
  r})^{1/3}+\frac{\hbar^{2}}{8m}\left[\frac{|\nabla\rho({\bf
      r})|^{2}}{\rho({\bf r})^{2}}-2\frac{|\nabla^{2}\rho({\bf
      r})|}{\rho({\bf r})}\right].
\end{array}
\label{mutfdw}
\end{equation}

Now we impose the statistical equilibrium $\mu({\bf r})=constant$ at the
interatomic surface $S'$ along the normal crossing direction, i.e. 
\begin{equation}
\nabla_{\bf n}\mu({\bf r})=0
\label{mu}
\end{equation}
\begin{equation}
\forall {\bf r} \in S'({\bf r}).
\label{sp}
\end{equation}
In order to understand a possible meaning of Bader's surface $S$ 
in a statistical
context, let us suppose that $S$ and $S'$ coincide, which means
that we suppose Bader's topological partitioning criterion is equivalent to a
statistical partitioning criterion of $\rho({\bf r})$, this will allow us 
to apply the properties of $S$ to Eq.\ref{mu} (of course $S$ in Eq.\ref{surface} and $S'$ in Eq.\ref{sp} are not
necessarily the same and will comment on this case later on); i.e.
taking into account that 
${\bf \nabla}\rho({\bf r})\cdot {\bf n}=0,\forall {\bf r} \in S({\bf r})$
we obtain
\begin{equation}
\nabla_{\bf n}\left[-\sum_{i=1,M}
\frac{Z_{i}}{|{\bf r}-{\bf R}_{i}|}+
\frac{1}{2}\int_{V}\frac{\rho({\bf r'})}{|{\bf r}-{\bf r}'|} d{\bf
  r}'+\frac{\hbar^{2}}{8m}\left[\frac{|\nabla\rho({\bf r})|^{2}}{\rho({\bf r})^{2}}-2\frac{|\nabla^{2}\rho({\bf r})|}{\rho({\bf r})}\right]\right]=0.
\label{force}
\end{equation}
Since $-\sum_{i=1,M}
\frac{Z_{i}}{|{\bf r}-{\bf R}_{i}|}+
\frac{1}{2}\int_{V}\frac{\rho({\bf r'})}{|{\bf r}-{\bf r}'|} d{\bf
  r}=\phi({\bf r})$ is the classical electrostatic potential,
Eq.\ref{force} expresses the condition that the
generalized force which takes into account non-local (and
non-classical)
effects must vanish at the surface of separation.
The interpretation of the above result is the same 
furnished in our previous work, with the exception that now we do not
simply have the classical electrostatic force but an additive term, as one
could have easily guessed {\it a priori}.
However what makes this result interesting is the fact that the term 
$\frac{\hbar^{2}}{8m}\left[\frac{|\nabla\rho({\bf r})|^{2}}{\rho({\bf
      r})^{2}}-2\frac{|\nabla^{2}\rho({\bf r})|}{\rho({\bf
      r})}\right]=Q({\bf r})$ represents the Bohm potential (see \cite{bohm4}). This consideration,
will allow us to interpret the results found so far in terms of
bohmian mechanics.
\section{Bader's theory and Bohmian Mechanics}
We supposed that $S$ and $S'$ are the same, i.e. Bader's interatomic
surface corresponds to a surface of separation between two statistical 
subsystems in equilibrium. $S$ (or $S'$) does not explicitly show the
nature of the bond at the separation, but only define the topological
domain of each atom. This means that this analysis furnishes 
one chemical information (defines atomic entities) but does not 
show how they are related to each other in chemical terms.
In their work \cite{bohm4} Levit and Sarfatti shows that Bohm's
quantum potential can give a direct interpretation, at least for atoms 
not beyond the third row as underlined by Hamilton \cite{bohm5,bohm6},
of bonding and reactivity in a multiatomic system. In the light of
what stated above let us consider the concept expressed by
Eq.\ref{force}, i.e. the electrostatic potential $\phi({\bf r})$
must be equal to minus Bohm potential $Q({\bf r})$ at every point of the 
surface $S$. This means that from the topology of $-\phi({\bf r})$ one
can work out the bonding nature and reactivity at each point of the
separation surface, i.e. along $S$ bonding and reactivity can be
determined in terms of classical electrostatic interaction. 

One can notice that in general, in case $S$ and $S'$ are not the same 
(i.e.$\nabla\rho({\bf r})\neq 0$), it is possible to say that 
along $S'$ bonding and reactivity are equivalently described by 
a quantum (Bohm potential) or a classical ($E_{TF}$) approach.  

Although the model used is highly simplified, as extensively discussed
in our previous work, the conclusions above are very interesting
because furnish basic indications about the nature of chemical properties in a
statistical context stimulating a further development at higher level
of theory.

 As stated in our previous work, rigorously speaking, the
force which acts at the surface of an atom is the Ehrenfest force
and the results obtained here are valid only within the approximation
done for exchange and density's inhomogeneity. At higher level of
approximation
for exchange and correlation, the physical picture may be very
different;
nevertheless starting from our results one could build a more complete 
and rigorous picture.
We have found also another connection with bohmian mechanics in terms
of single particle dynamics. In this case we have to consider the 
quantum wavefunction to be imaginary and in polar form $\psi({\bf r})=R({\bf
  r},t)e^{iS({\bf r},t)}/\hbar$, or the more familiar $\psi({\bf
  r})=(\rho({\bf r},t)^{-1/2})e^{iS({\bf r},t)}/\hbar$ where $\rho({\bf r},t)$ is the one
particle time dependent electron
density as considered so far.
The point particles of mass $m$ in the configuration space will be
governed by the following dynamics:
\begin{equation}
v_{n}=\frac{\hbar}{m}Im\frac{\nabla_{n}\psi}{\psi}
\label{velocity}
\end{equation}
where $v_{n}$ is the velocity of the particle along the direction $\bf 
n$. Applying the Shr\"{o}edinger equation and considering the
stationary case, using the property $\nabla\cdot v_{n}=\frac{\partial
  v_{n}}{\partial{n}}$ where $\frac{\partial }{\partial{n}}$ is the derivative with respect
to the direction normal to the surface $S$, from equation (2) in Ref.\cite{bohm2} we obtain:
\begin{equation}
\frac{\partial v_{n}^{2}}{\partial{n}}=-\nabla_{n}\left(\frac{1}{m}\phi({\bf r})+Q({\bf
    r})\right)
\label{bd}
\end{equation}
($\nabla_{n}={\bf n}\cdot\nabla$, and ${\bf n}$ as specified before is 
the vector normal to $S$ at each point).
Considering $S$ and $S'$ coinciding, from Eq.\ref{force} and
Eq.\ref{bd} one obtains:
\begin{equation}
v_{n}\frac{\partial v_{n}}{\partial{n}}=0
\label{kin}
\end{equation}
Eq.\ref{kin} is satisfied either in case the velocity of the
particle normal to $S$ is zero or in case the flux of kinetic energy
is zero.
In the first case we have a natural interpretation on interatomic
surface,
i.e. the particle does not cross the border and so belongs to a
particular topological domain, in the second case, this
expresses the fact that the flux of classical kinetic energy in the
Thomas-Fermi formulation $\frac{\nabla\rho}{\rho^{1/3}}$ and the flux
of the quantum kinetic energy $v_{n}\frac{\partial
  v_{n}}{\partial{n}}$
through the surface $S$ are both zero. These properties of $S$ are
very interesting and, as the properties previously shown, 
stimulate further investigation of the
classical-quantum connections of $S$. However one must notice that
although this latter result is intriguing, it applies only to complex
wavefunctions; this means that 
for practical application in current research, where 
ground state calculations are mainly done, this result cannot be used
unless particular cases are investigated (e.g. systems in magnetic
fields).

\section{Discussion and Conclusions}
We analyzed the statistical meaning of interatomic surface in Bader's 
formulation from a
statistical point of view within the approximation of
Thomas-Fermi-Dirac-Wiezsacker. It emerged an interesting connection
with bohmian mechanics in interpreting properties of 
points at the surface $S$ in terms of bonding and reactivity.
It must be also clear that this analysis is valid in the limit of the 
approximations done, more sophisticated models could lead to different
results and certainly not to a direct interpretation in terms of
bohmian mechanics. However it would be interesting to 
understand why and how the modification of  properties of $S$ occur at 
higher level of theory using our previous analysis as a starting point.
Another important problem is to understand how $S$ and $S'$ are
connected for real cases; our analysis was based on the hypothesis that they 
do coincide, so that we could comment on the consistency of Bader's
surface with a statistical definition but this hypothesis is not
obvious and if they differ it would be interesting to understand at
which level of approximation they do coincide. 
Since Bader's procedure to obtain $S$ is based on a rigorous method within 
quantum mechanics, the exact relation between $S$ and
$S'$ at different levels of statistical theory could lead to
clearer formulation of quantum properties in statistical terms with
obvious advantages for practical applications.
In conclusion, we think that our analysis is important for a
formulation in statistical terms of the chemical properties at
atomistic level and because represents a stimulating challenge for
further developments.

\end{document}